\documentclass[manuscript]{aastex631}

\begin{document}

\title{A Distance Measurement to M33 Using Optical Photometry of Mira Variables}

\correspondingauthor{Jia-Yu Ou}
\email{m1039004@gm.astro.ncu.edu.tw}

\author[0000-0002-6928-2240]{Jia-Yu Ou}
\affil{Graduate Institute of Astronomy, National Central University, 300 Jhongda Road, 32001 Jhongli, Taiwan}
\author[0000-0001-8771-7554]{Chow-Choong Ngeow}
\affil{Graduate Institute of Astronomy, National Central University, 300 Jhongda Road, 32001 Jhongli, Taiwan}

\author[0000-0001-6147-3360]{Anupam Bhardwaj}
\affil{INAF-Osservatorio astronomico di Capodimonte, Via Moiariello 16, 80131 Napoli, Italy}

\author[0000-0002-3168-0139]{Matthew J. Graham}
\affiliation{Division of Physics, Mathematics, and Astronomy, California Institute of Technology, Pasadena, CA 91125, USA}

\author[0000-0003-2451-5482]{Russ R. Laher}
\affiliation{IPAC, California Institute of Technology, 1200 E. California Blvd, Pasadena, CA 91125, USA}

\author[0000-0002-8532-9395]{Frank J. Masci}
\affiliation{IPAC, California Institute of Technology, 1200 E. California Blvd, Pasadena, CA 91125, USA}
             
\author[0000-0002-0387-370X]{Reed Riddle}
\affiliation{Caltech Optical Observatories, California Institute of Technology, Pasadena, CA 91125, USA}

\begin{abstract}

We present a systematic analysis to determine and improve the pulsation periods of 1637 known long-period Mira variables in M33 using $gri$-band light curves spanning $\sim18$~years from several surveys, including M33 variability survey, Panoramic Survey Telescope and Rapid Response System, Palomar Transient Factory (PTF), intermediate PTF, and Zwicky Transient Facility. Based on these collections of light curves, we found that optical band light curves that are as complete as possible are crucial to determine the periods of distant Miras. We demonstrated that the machine learning techniques can be used to classify Miras into O-rich and C-rich based on the $(J-K_s)$ period--color plane. Finally, We derived the distance modulus to M33 using O-rich Miras at maximum light together with our improved periods as $24.67 \pm 0.06$~mag, which is in good agreement with the recommended value given in the literature. 
\end{abstract}
%We conducted a multiband period analysis and a photometric investigation for Mira variable stars in the local galaxy M33. We collected over 15 years of $gri$-band data from several surveys, including M33 variability survey, Panoramic Survey Telescope and Rapid Response System (Pan-STARRS), Palomar Transient Factory (PTF), intermediate PTF (iPTF), and Zwicky Transient Facility (ZTF). We revealed differences between Miras in period--magnitude--color and period--color diagrams, and we subsequently used machine learning algorithms to categorize Mira subclasses based period--color differences. Finally, we determined the period-luminosity relationships at both maximum light and mean light for M33 Mira variable stars. The derived distance modulus to M33 using Mira variable stars is $24.67 \pm 0.05$~mag and $24.75 \pm 0.05$~mag at maximum and mean light, respectively.

%\keywords{}

\section{Introduction} \label{sec:intro}
%In 1596, David Fabricius discovered the first Mira  ($o$ Ceti), and since then, research has been conducted on Mira variable stars (hereafter Miras). Miras are red giants located on the asymptotic giant branch (AGB); AGB stars, which are at the last stage of evolution, are of low or intermediate mass, and their mass ranges from ~0.8 to 8.0 times the mass of the Sun \citep{2012ApSS.341..123W}. 

Mira variable stars (hereafter Miras) are red giants located on the asymptotic giant branch (AGB) at their late stage of evolution. Miras are long-period variable (LPV) stars, with periods ranging from hundreds to thousands of days, and they have large amplitude variations in the optical and near-infrared (NIR) bands \citep[for examples, see][]{2009AcA....59..239S,2010ApJ...723.1195R, 2012ApSS.341..123W}. Miras can be categorized as oxygen-rich (O-rich) or carbon-rich (C-rich) stars according to the nature of molecules that dominate their spectra; this  depends on the C/O ratio presents in their atmosphere \citep{1960ApJ...131..385M,2001A&A...377..945C,2010ApJ...723.1195R}. Alternatively, division of Miras into 
 O-rich or C-rich candidates can also be done by using photometric data when the spectroscopic data is lacking. For examples, Miras were categorized on the basis of period and Wesenheit indices $W_I$ in \citealt{2005AcA....55..331S}. On the other hand, \citealt{2009AcA....59..239S} found that both categories of stars can be separated using a $(V - I)$ versus $(J - K_s)$ and $W_{JKs}$ versus $W_I$ diagram. In \citealt{2017AJ....153..170Y} and \citealt{2018AJ....156..112Y}, the authors also divided stars into subclasses by using a $(J - K_s)$ versus ($H - K_s)$ approach. In \citealt{2018AA...616L..13L}, the authors used the $(W_{RP} - W_{K_s})$ versus $K_s$ to divide the O-rich and C-rich AGB stars based on the Gaia and 2MASS data. O-rich and C-rich Miras were also found to belong to two distributions in a period and $(J - K_s)$ diagram \citep{2021ApJ...919...99I}.

\begin{figure*}
    \centering
    \epsscale{1.15}
    \plotone{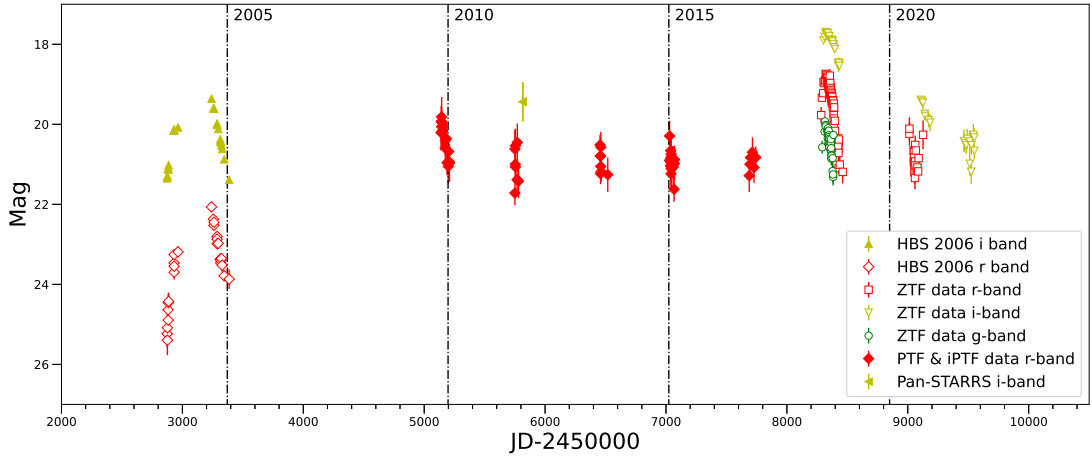}
    \caption{The $gri$-band light curves for HBS 2006--40671, a Mira identified in the M33 variability survey \citep{2006MNRAS.371.1405H}, subsequently spectroscopically confirmed by \citet{2011MNRAS.413.1797B}. The green, red, and yellow symbols represent the $gri$-band light curves, respectively.}
    \label{40671_LC}
\end{figure*}

Since the first period--luminosity (PL, also known as the Leavitt Law) relation for Miras was published in \citet{1981Natur.291..303G} using NIR data, a number of researchers have attempted to derive and calibrate the Mira P--L relations \citep[e.g., see][]{1984MNRAS.211P..51F,1989MNRAS.241..375F,1997MNRAS.289..428K,2008MNRAS.386..313W,2017AJ.154..149Y,2019ApJ...884...20B,2021ApJ...919...99I,2022arXiv220300896O}. The dispersion of the Mira P--L relation at maximum light was found to be smaller than that of their counterparts at mean light \citep{1997MNRAS.289..428K,2019ApJ...884...20B,2022arXiv220300896O}.

Recently, we have derived and calibrated the $I$-band PL relations for Miras located in the Magellanic Clouds \citep{2022arXiv220300896O}. Hence, the main goal of this work is to test the applicability of our PL relations in measuring distances to nearby galaxies. We selected M33 for our test because there is a sizable sample of Miras found in M33 \citep{2018AJ....156..112Y}, M33 is close enough such that light curves for these Miras can be retrieved from archives (Section \ref{sec2}), and there are numerous independent distance measurements to M33 so we can compare our derived distances with these independent measurements. The sample of 1781 M33 Miras compiled in \citet{2018AJ....156..112Y} was classified into 88 C-rich Miras, 1265 O-rich Miras, and 428 of unknown type. Multiple reasons were accounted for the unknown type, as described in \citet{2018AJ....156..112Y} and will not be repeated here. Using independent optical band light curves we collected in Section \ref{sec2}, we re-determined the periods for this sample of M33 Miras, as well as their magnitudes at mean and maximum light, and reclassified the Miras with unknown type using a machine learning approach. The results of this analysis are presented in Section \ref{sec3}.  Since C-rich Miras could exhibit a significant long-term variation in their light curves \citep{2021ApJ...919...99I,2022arXiv220300896O} which will affect the result of derived PL relation, we only use O-rich Miras (including those being reclassified) to determine the distance to M33 using our derived PL relation \citep{2022arXiv220300896O} and compared to other distance measurements in Section \ref{sec4}, followed by the conclusion of this work in Section \ref{sec5}.  

%The aims of this study were to (a) analyze multiband periods using an M33 distance modulus and optical-band data from several surveys and (b) use machine learning technology to construct a classifier that can divide  O- and C-rich Miras between a period and J -- Ks. 

%The rest of this paper is organized as follows. Section 2 describes the data used in this study. Section 3 describes the period analysis results, color magnitude diagram, period color diagram, and machine learning results. In Section 3, we discuss the PL relationship with regard to maximum light time and mean light time. The results are summarized in Section 4.

\section{Archival Light Curves}\label{sec2}

We collected the long-term (2003 to 2021) optical $gri$-band light curves for the 1781 M33 Miras compiled in \citet{2018AJ....156..112Y} from various sources. We cross-matched this sample of Miras to the variable sources detected in the M33 variability survey \citep[][hereafter HBS 2006]{2006MNRAS.371.1405H}, a time-series $gri$-band survey carried with the MegaCam mounted on the Canada–France–Hawaii Telescope (CFHT) from 2003 to 2005 (for 27 nights). Sparse $gri$-band light curves data were also collected (whenever available) from the Panoramic Survey Telescope and Rapid Response System \cite[Pan-STARRS,][]{2010SPIE.7733E..0EK,2016arXiv161205560C} Data Release 2. We then extracted the $R_{PTF}$-band light curves for the M33 Miras from the Palomar Transient Factory \cite[PTF,][]{2009PASP..121.1395L,2009PASP..121.1334R} and the intermediate PTF \cite[iPTF,][]{2013ATel.4807....1K}. These $R_{PTF}$-band light curves were calibrated to the $r$-band using the Pan-STARRS photometric catalog.  Specifically, we performed differential photometry by selecting a number of suitable reference stars around each Miras, where the $r$-band magnitude for these reference stars are available from the Pan-STARRS photometric catalog. Together, the Pan-STARRS and the PTF/iPTF light curves spanned from 2009 to 2017. Finally, the $gri$-band light curves data after 2017 to 2021 were collected from the 
Zwicky Transient Facility \cite[ZTF,][]{2019PASP..131a8002B,dec20,2019PASP..131g8001G,2019PASP..131a8003M} Data Release 10 and the ZTF collaboration survey data.\footnote{All ZTF data, including the collaboration surveys data \citep[described further in][]{2019PASP..131a8002B}, were processed using the same dedicated ZTF reduction pipeline as described in \citet{2019PASP..131a8003M}.} All-together, we collected $gri$-band light curves data for 1367 M33 Miras (see Table \ref{data}). An example of the collected light curve is shown in Figure \ref{40671_LC}. We noted that since the telescopes used in Pan-STARRS and PTF/iPTF/ZTF have an aperture of 1.8-m and 1.2-m, respectively, limiting magnitudes from these surveys are around $\sim 19.5$~mag to $\sim 21.5$~mag. In contrast the deep CFHT observations for the M33 variability survey can reach to a depth of $\sim 25$~mag in $r$-band.

\begin{figure}
    \epsscale{1.1}
    \plotone{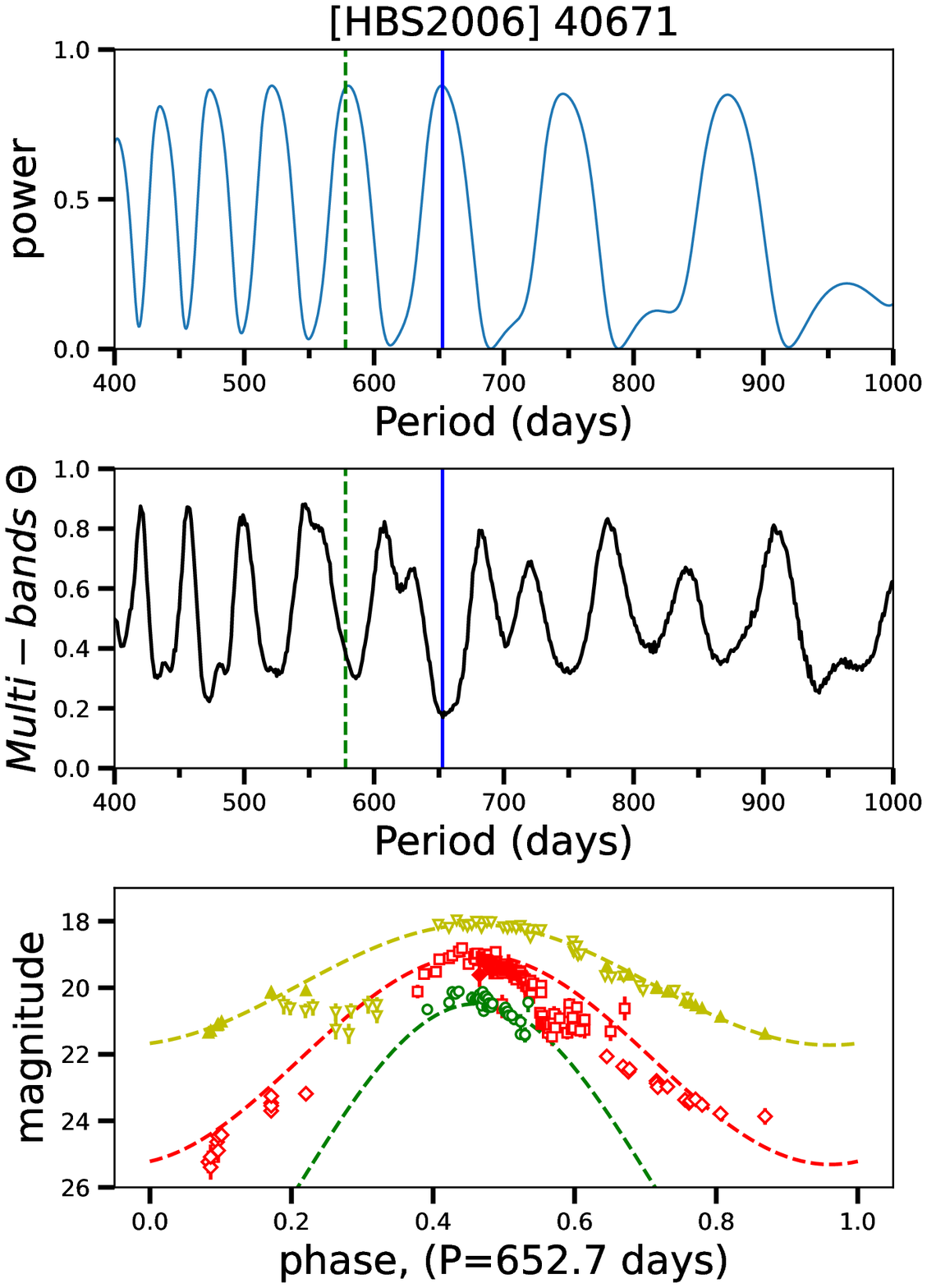}
    \caption{Top panel: Multi-band LS periodogram for M33 Mira HBS 2006--40671.  The vertical blue solid line represents the highest peak which corresponding to a period of $652.7 \pm 23.3$~days. The green dashed line is the identified second highest peak at a period of $578 \pm 32$ days. Due to long period nature of Miras and time-sampling of ground-based observations, aliasing that causing multiple peaks is an unavoidable issue for LS-based periodogram. Middle panel: The multi-band PDM periodogram for the same M33 Mira, where the locations of the blue solid and green dashed lines are same as the top panel. Minimum value of multi-band $\Theta$ represents the most probably period.  The multi-band PDM was used to cross-check the LS period, especially when multiple peaks are presented in the LS periodogram. Botton panel: Phased $gri$-band light curve folded with the determined period. The symbols are same as in Figure \ref{40671_LC}. The dashed curves are the fitted sinusoidal function to the light curve data. }
    \label{40671_LS}
\end{figure}

\begin{deluxetable}{cccccc}
    \label{data}
    \tabletypesize{\footnotesize}
    \tablecaption{Light Curve Data of Miras in M33.}
    \tablecolumns{6}
    \tablewidth{0pt}
    \tablehead{
	\colhead{ID} & \colhead{Filter} & \colhead{MJD} & \colhead{Mag} & \colhead{Error} & \colhead{Source} 
          }
\startdata
01321450+3019349 &i &59140.343576 &21.09 &0.30  &ZTF   \\   
01321450+3019349 &i &59460.492639 &20.13 &0.10  &ZTF   \\ 
01321450+3019349 &i &59469.454479 &20.56 &0.20  &ZTF\\
$\cdots$ &$\cdots$ &$\cdots$& $\cdots$ & $\cdots$ &$\cdots$ \\
\enddata
\tablecomments{The entire Table is published in its entirety in the machine-readable format. A portion is shown here for guidance regarding its form and content.}
\end{deluxetable}

\section{Analysis and Results}\label{sec3}

\subsection{Periods Determination}\label{sec31}

Since Miras exhibit a large amplitude variation in the optical bands, it is possible that for Miras located in a distant galaxy, only a portion of the optical-band light curve (i.e. around the maximum light) brighter than the limiting magnitude of a given survey can be detected. This is indeed seen in the light curves for M33 Miras collected from Pan-STARRS, PTF/iPTF, and ZTF, as demonstrated in Figure \ref{40671_LC}. On the other band, the deep HBS 2006 observations can sample the  full amplitude light curves, including the portion of the light curves around the minimum light. These combinations provide us an opportunity to test the period determination for distant Miras when only a portion of the light curves above a given detection limit is available.

We first determined the periods for our sample of M33 Miras using the multi-band Lomb--Scargle (LS) periodogram, described in \citet{2015ApJ...812...18V} and implemented in the {\tt gatspy} package, on the full set of $gri$-band light curves (whenever available).  Errors on the determined periods were estimated based on bootstrap resampling method.  We then checked the LS periods using a multi-band Phase Dispersion Minimization (PDM) periodogram  developed in \citet{2021ApJ...911...51L}. If both periods agree (e.g. the periods are within $10\%$ of each others) we adopted the LS periods, else we visually inspected the phased light curves and selected the period that resulted a smoother light curve. Upper panel of Figure \ref{40671_LS} presents an example of the LS periodogram for HBS 2006--40671 (see Figure \ref{40671_LC} for the observed light curves), at which a period of $652.7 \pm 23.3$~days was identified.\footnote{Using a PDM  \citep{1965ApJS...11..216L} technique, \citet{2011MNRAS.413.1797B} found that this Mira has a period of 665~days. However, a shorter period of $578\pm32$~days was identified by \citet{2017AJ....153..170Y} using a semi-parametric periodogram technique. Furthermore, the multi-band periodogram applied in \citet{2018AJ....156..112Y} found 426 and 654~days as the primary and secondary periods, respectively.}  The periodogram from the multi-band PDM, as presented in the middle panel of Figure \ref{40671_LS}, also picked up the same period as the LS periodogram. Even though the LS periodograms could have multiple peaks with similar heights, we applied both multi-band LS and PDM methods for cross-check and validations to ensure the most probable periods were selected.

\begin{figure*}
    \epsscale{1.1}
    \plotone{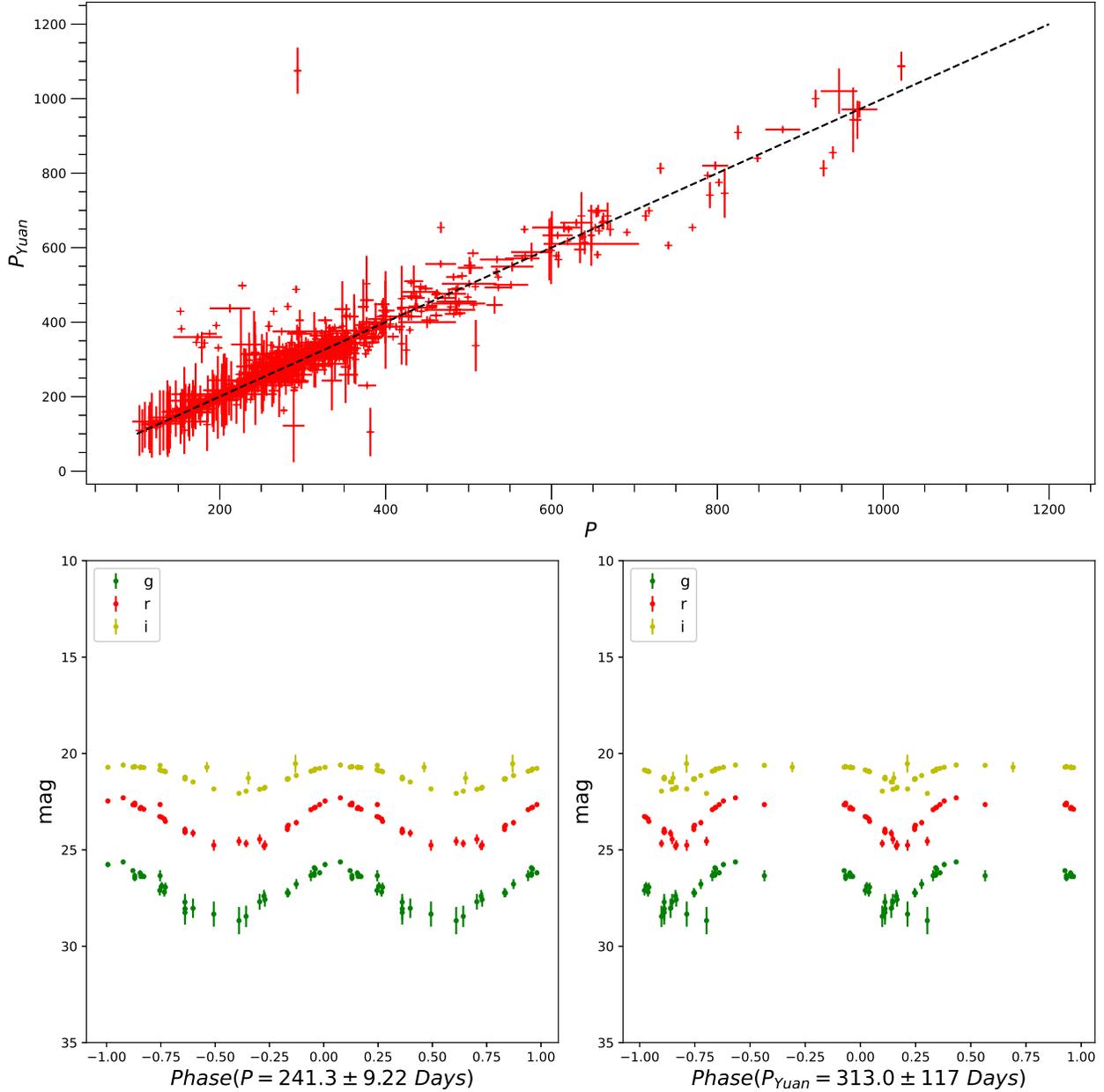}
    \caption{Top panel: Comparison of the periods determined from this work ($P$) and the periods published in \citet[][$P_{yuan}$]{2018AJ....156..112Y}. The dashed line represents the $1:1$ relation. Note that both light curves data and period-search methodologies are totally independent between our work and \citet{2018AJ....156..112Y}. Bottom panels: An example of the $gri$-band light curves folded with periods determined in this work (bottom-left panel) and in \citet[][bottom-right panel]{2018AJ....156..112Y}.}
    \label{PP}
\end{figure*}

\begin{figure*}
    \plotone{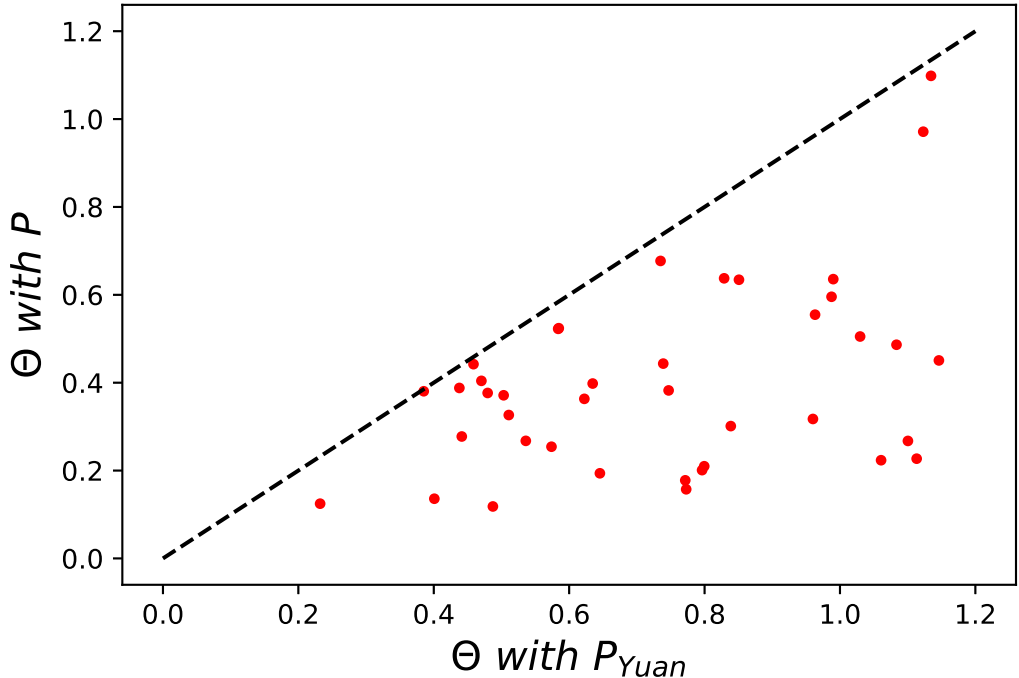}
    \caption{Phase dispersions $\Theta$ of the light curves based on the folded period of $P$ and $P_{\mathrm{Yuan}}$ for the $\sim4\%$ of Mira showing discrepant periods.The dashed line represents the $1:1$ relation.}
    \label{pd}
\end{figure*}

Our determined periods are given in Table \ref{datasummary}. In general, our determined periods agreed with periods presented in \citet{2018AJ....156..112Y}, as evident in the upper panel of Figure \ref{PP},  validating our period-search method. For this sample of M33 Miras, only $\sim 4\%$ of the Miras display a significant difference in the periods we found here and in \citet{2018AJ....156..112Y}, an example is presented in the bottom panel of Figure \ref{PP}. Figure \ref{pd} shows that for these $\sim 4\%$ of Mira, the overall phase dispersions for light curves folded with our determined periods are generally smaller than those using the periods from \citet{2018AJ....156..112Y}.

%For discrepant periods, we visually inspected phased light curves folded with both periods, 
%{\bf and only less than 5\% period were significant differences (an example is given in the bottom panels of Figure \ref{PP} and Figure \ref{pd}).}
%and found that our determined periods can fold the light curves better (an example is given in the bottom panels of Figure \ref{40671_LS}).  

\begin{figure*}
    \epsscale{1.17}
    \plottwo{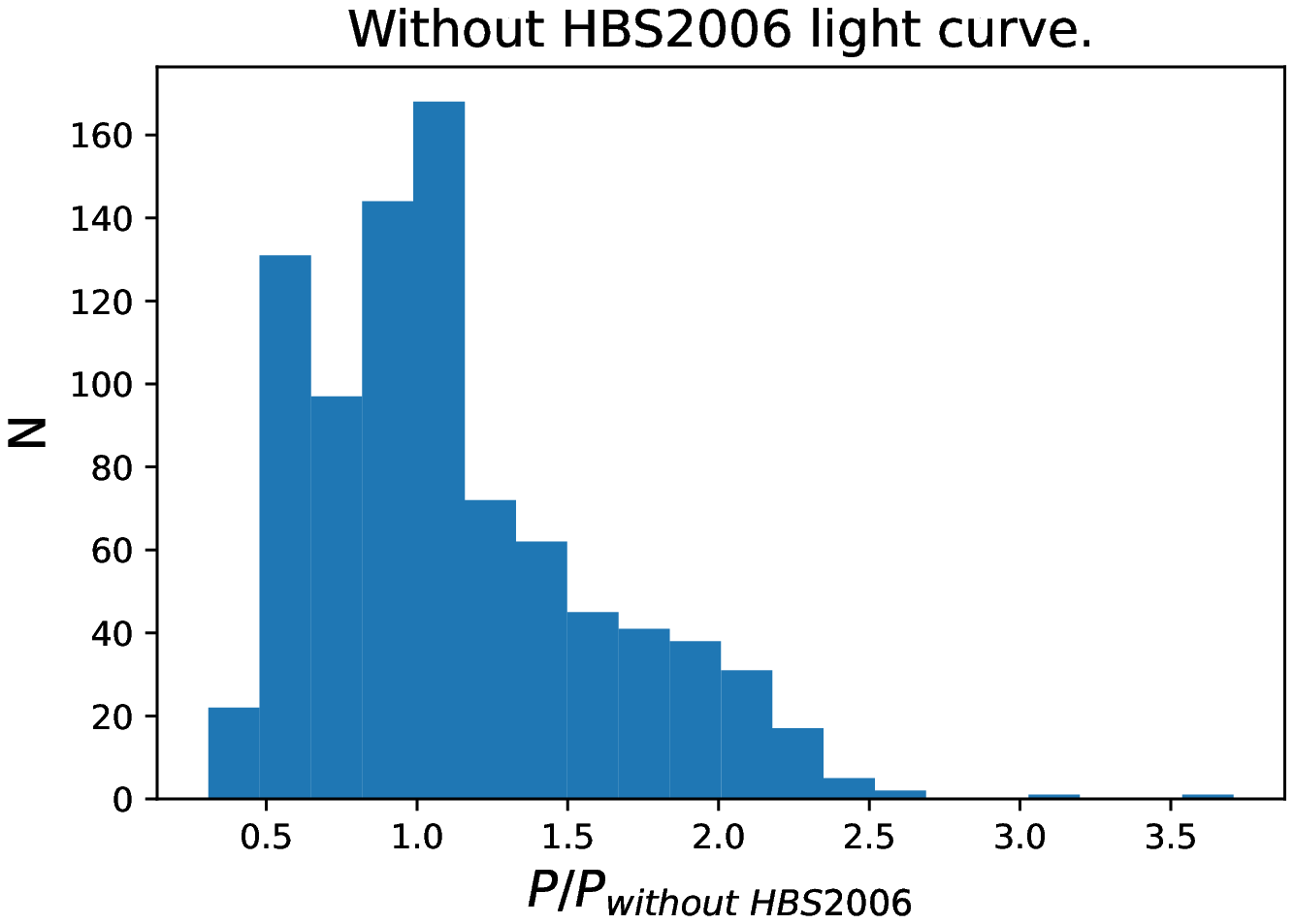}{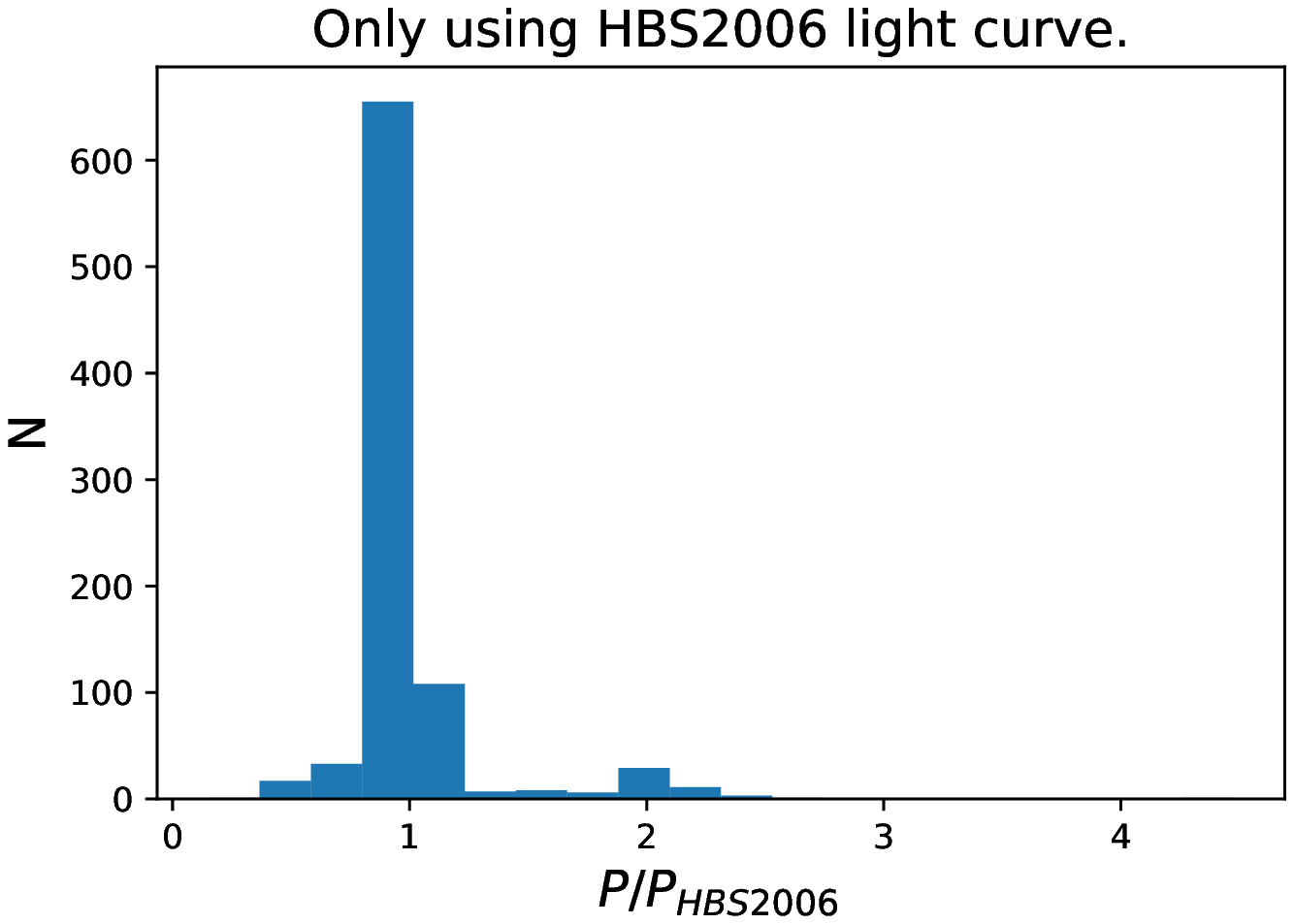}
    \caption{Comparisons of the periods when the portions of the light curves from the deep HBS2006 observations were excluded (left panel) and only using the observed light curves from HBS2006 (right panel). The abscissa on both histograms represent the period ratios, where $P$ (same as in the upper panel of Figure \ref{PP}) are the determined periods using the full light curves.}
    \label{Period_result}
\end{figure*}

\begin{figure*}
    \begin{tabular}{ccc}
    \includegraphics[width=0.3\columnwidth]{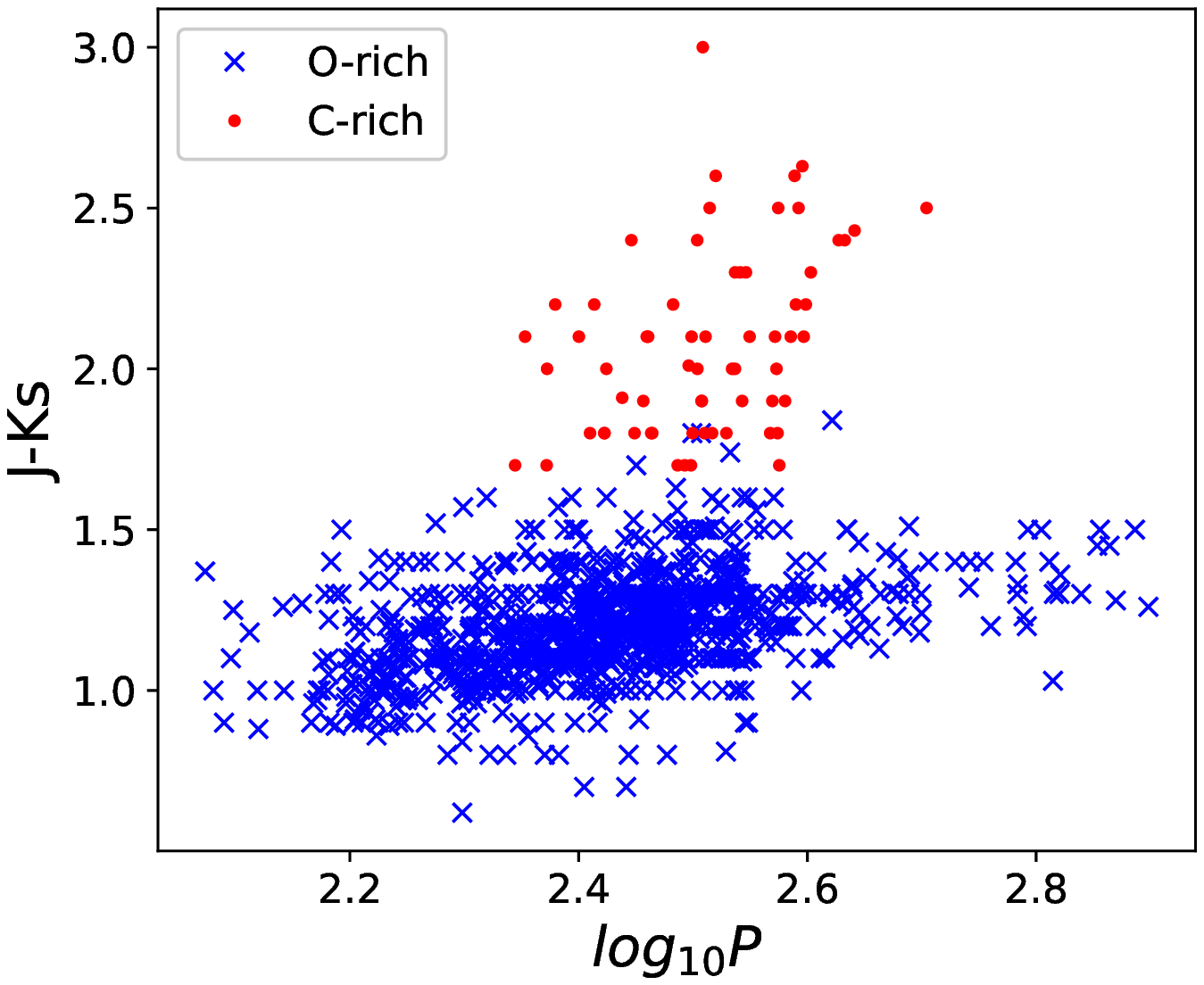} & \includegraphics[width=0.3\columnwidth]{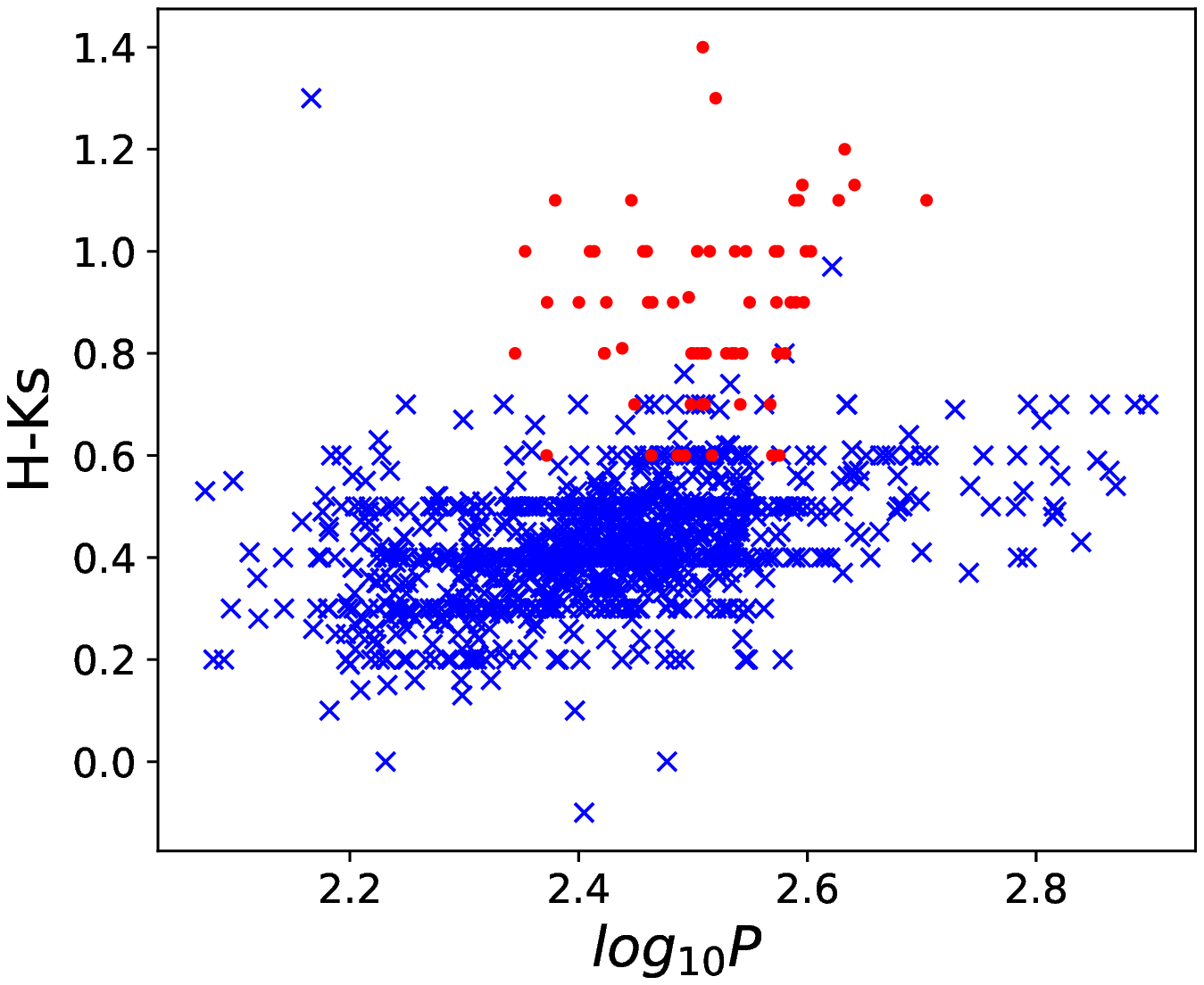} & \includegraphics[width=0.3\columnwidth]{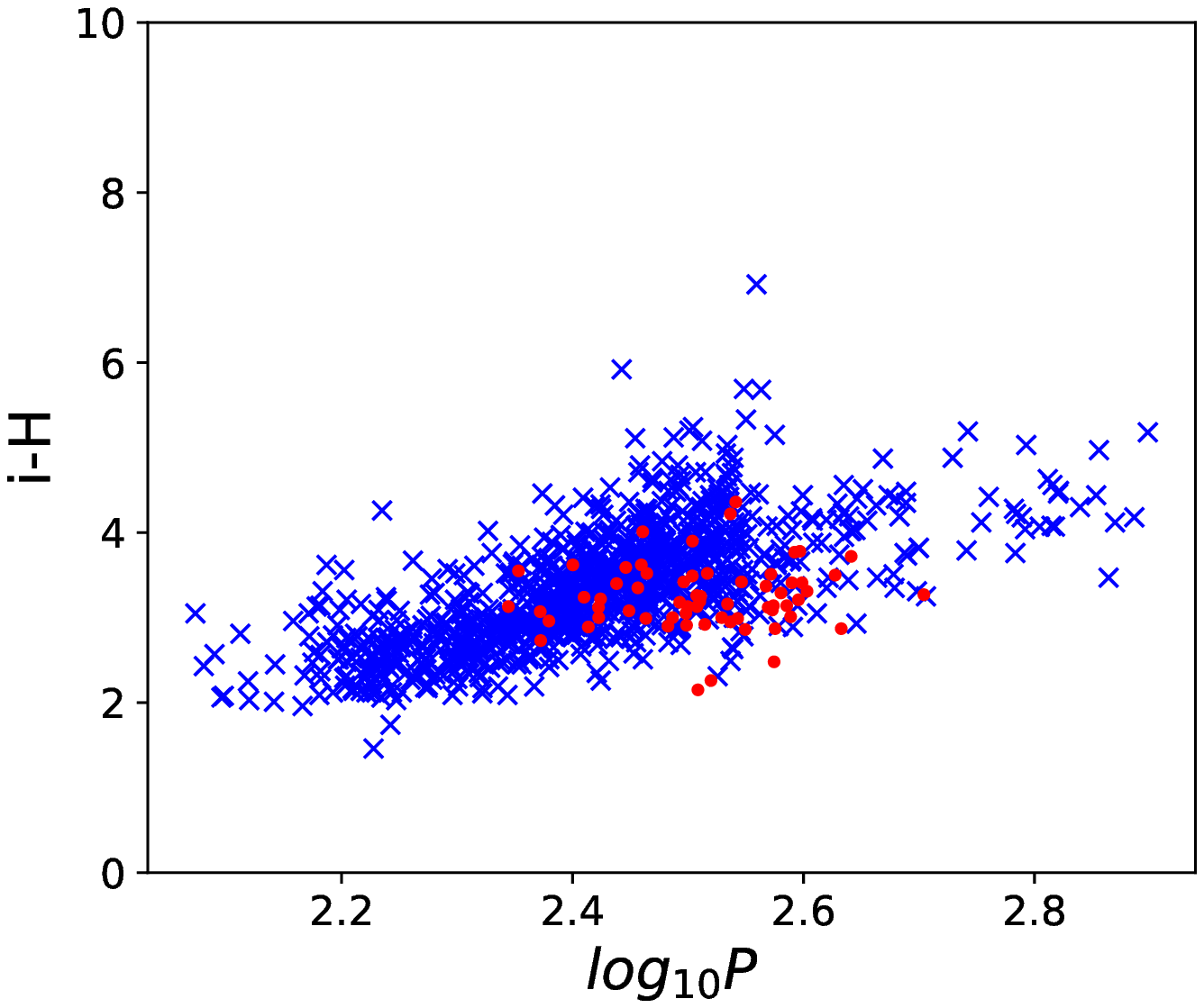} 
    \end{tabular}
    \caption{The period--color relations for M33 Miras, where the classification of O-rich (blue crosses) and C-rich (red dots) Miras were provided by \citet{2018AJ....156..112Y}.}
    \label{PColor}
\end{figure*}

The deep but short duration (probably cover $\sim 1$ to $\sim 2$ pulsation cycles of Miras) of CFHT observations from HBS 2006 and the shallow but longer duration (covering more than two pulsation cycles) from the Pan-STARRS/PTF/iPTF/ZTF surveys represent two typical cases encountered in dedicated deep time-series observations (such as targeting a particular galaxy) or synoptic sky surveys. We tested two scenarios on period-search by excluding the deep HBS 2006 light curves (and only using light curves from Pan-STARRS/PTF/iPTF/ZTF) and only using the HBS 2006 light curves. Comparisons of the periods found in these two scenarios to the assumed true period $P$ using the full HBS 2006 and Pan-STARRS/PTF/iPTF/ZTF light curves are presented in Figure \ref{Period_result}. Our test results show that to recover the periods, it is important to sample the full amplitude light curves for Miras rather than only sampling a portion of the light curves around maximum light covering few pulsation cycles.

\subsection{Magnitudes at Mean and Maximum Light}\label{sec32}

%\begin{figure*}
%    \epsscale{1.1}
%    \plottwo{CMD_JKs_O-rich.eps}{CMD_JKs_C-rich.eps}
%    \plottwo{CMD_iJ_O-rich.eps}{CMD_iJ_C-rich.eps}
%    \plottwo{CMD_rH_O-rich.eps}{CMD_rH_C-rich.eps}
%    \caption{CMD at which one of the magnitude included the $JHK_s$ filters. The left panels show O-rich Miras, and the right panels present C-rich Miras. The color bars represent pulsation periods.}
%    \label{NIRCMD}
%\end{figure*}

%\begin{figure*}
%    \epsscale{1.17}
%    \plottwo{CMD_ri_O-rich.eps}{CMD_ri_C-rich.eps}
%    \plottwo{CMD_gi_O-rich.eps}{CMD_gi_C-rich.eps}
%    \caption{Similar to Figure \ref{NIRCMD}, except the photometry only invloved $gri$ filters.}
%    \label{optCMD}
%\end{figure*}

Similar to \citet{2018AJ....156..112Y},  we fit a sinusoidal function in the form of $m = m_0 + A_1 \sin( \omega t + \phi_1) + A_2 \sin( 2\omega t + \phi_2)$, where $\omega = 2\pi / P$, to the folded $gri$-band light curves using the periods determined in previous subsection (see the bottom panel of Figure \ref{40671_LS} for an example). We then adopted $m_0$ as the magnitudes at mean light, $\langle m \rangle = m_0$ (in magnitude scale), from the fitted light curves, where $m=\{g,\ r,\, i\}$. We have also determined the magnitudes at maximum light, $m_{MAX}$, based on the same fitted light curves. These magnitudes are provided in Table \ref{datasummary}. Since not all of the Miras have well-sampled light curves in the $gr$-band, we caution that the $gr$-band magnitudes at mean light may not be reliable. On the other hand, $\langle i \rangle$ and $m_{MAX}$ in all three bands do not have such a problem. Extinction corrections on these magnitudes were conducted using a 3D dust map \citep[][by  converting the returned extinction values to $E(B-V)$ as listed in the last column of Table \ref{datasummary}]{2018MNRAS.478..651G} and adopting an average $R_V$ value of 3.39 \citep{2022ApJS..260...41W} for M33 in accordance with the \citet{1989ApJ...345..245C} reddening law. We have also compiled the NIR $JHK_s$ mean magnitudes for these M33 Miras from \citet{2018AJ....156..112Y}. Extinction corrections on these NIR mean magnitudes were done using $A_J = 0.03$~mag, $A_H = 0.02$~mag, and $A_{Ks} = 0.01$~mag \citep{2018AJ....156..112Y}.

%Figure \ref{NIRCMD} shows the CMD for O-rich (left panels) and C-rich (right panels) Miras, as classified in \citep{2018AJ....156..112Y}, at which the photometry included one or both of the $JHK_s$ filters. For the O-rich Miras, it is clear that longer period O-rich Miras tend to have redder colors and brighter magnitudes. In contrast, CMD constructed from optical $gri$ filters do not display such a trend (Figure \ref{optCMD}). The trends are not obvious for C-rich Miras due to a much smaller sample size than the O-rich Miras.

%On the basis of an NIR color magnitude diagram, as presented in Figure \ref{NIRCMD}, in the longer period, Mira variable stars were distributed in the redder and brighter parts. 

%{\bf NOT CLEAR WHAT THESE MEAN} The shorter-period Miras were distributed in the bluer and fainter parts. This is because the CMD distribution in the optical band does not have a wide distribution in shorter-period Miras, meaning that shorter-period Miras may be more affected by reductions in dust. However, the longer-period Mira variable stars still had a color--magnitude distribution similar to that in the NIR.

\subsection{Reclassification of M33 Miras with Unknown Type}\label{sec33}

%Similar to \citet{2018AJ....156..112Y}, we fit a sinusoidal function to the folded $gri$-band light curves using the periods determined in previous subsection (see the bottom panel of Figure \ref{40671_LS} for an example). We then determined the mean magnitudes, $\langle m \rangle$, and the magnitudes at maximum light, $m_{MAX}$, from the fitted light curves, where $m=\{g,\ r,\, i\}$. 

In accordance with the results presented in \citet{2021ApJ...919...99I}, we examined the period--color (PC) relations for the M33 Miras in various colors. Figure \ref{PColor} reveals that the O-rich and C-rich Miras exhibit different distributions in the $(J-K_s)$ and $(H-K_s)$ PC relations, but their distributions were similar in the $(i-H)$ PC relations. This implies that the colors of O-rich Miras is markedly different from that of C-rich Miras in NIR but not in the optical. That is caused by the C-rich Miras having higher abundance of circumstellar dust, which can significantly influence the near infrared radiation \citep{2021ApJS..257...23I,2022arXiv220300896O}. Figure \ref{PColor} also shows that the O-rich and C-rich M33 Miras can be well separated in the $(J-K_s)$ PC relation.

\begin{figure}
    \epsscale{0.7}
    \plotone{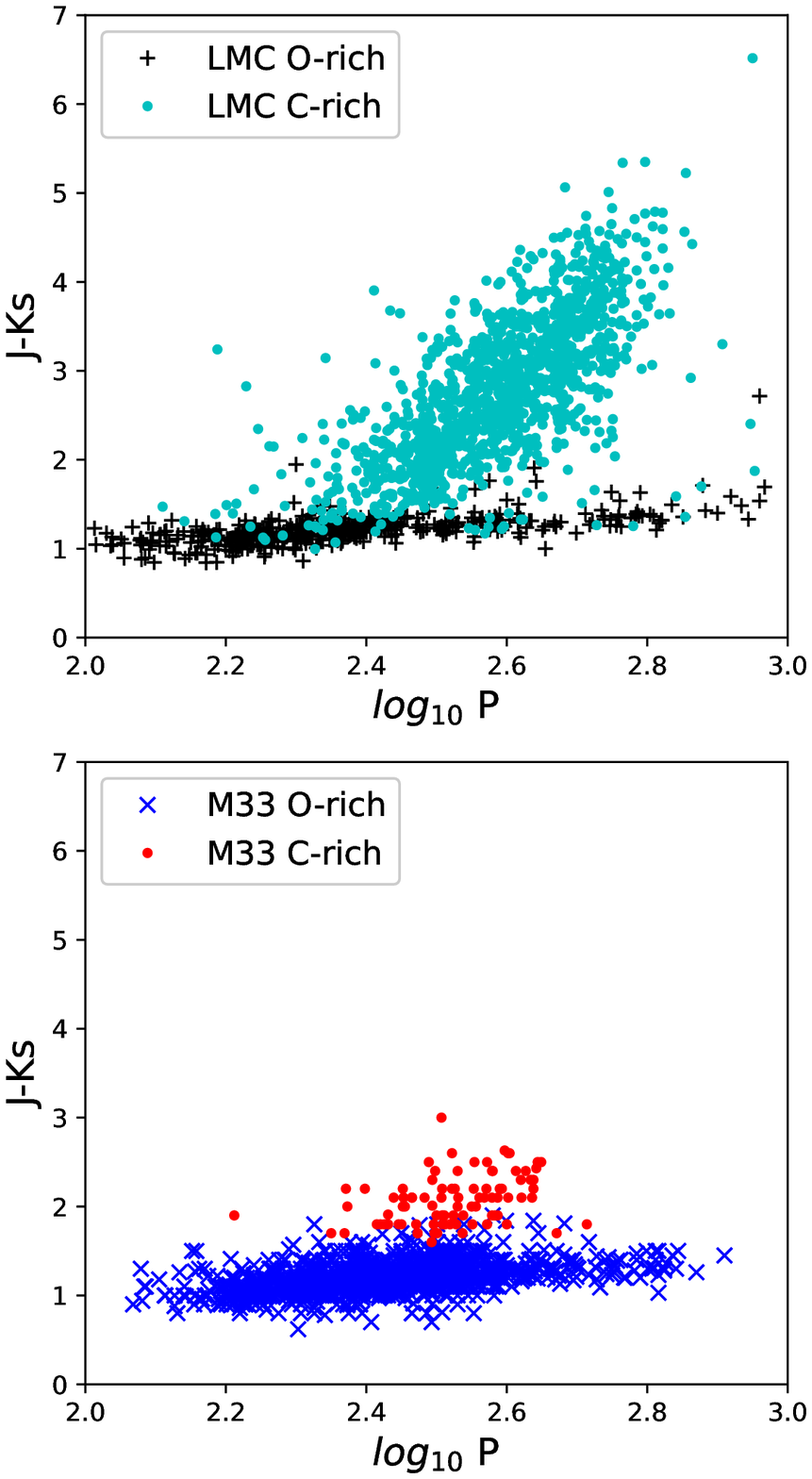}
    \caption{The $(J-K_s)$ PC relations for the samples of Miras in LMC (top panel0 and M33 (bottom panel).}
    \label{traning}
\end{figure}

The sample of Miras in the Large Magellanic Cloud \citep[LMC; as compiled in][]{2022arXiv220300896O}, the O-rich and C-rich Miras have different distributions in the $(J-K_s)$ PC relation, as indicated in Figure \ref{traning}. This suggested the O-rich and C-rich Miras can be classified in the $(J-K_s)$ PC plane via machine learning (ML) techniques. We tested four ML classifiers, namely the perceptron learning algorithm (PLA), the logistic regression algorithm (LRA), the K-nearest neighbor (KNN) algorithm, and the support vector machine (SVM) algorithm.
%on a combined sample of 3001 LMC and M33 Mira variables.

%we collected LMC NIR spectroscopy data by using 2MASS, and the period was obtained from \citep{2022arXiv220300896O}, as indicated in Figure \ref{traning}. The two Mira subtypes had different period-versus-color distributions, and we used machine learning to categorize these Mira subtypes. We tested four classifiers, namely the perceptron learning algorithm (PLA), K-nearest neighbor (KNN) algorithm, logistic regression algorithm (LRA), and support vector machine (SVM) algorithm, on 3001 Mira samples.

In the PLA, the training data were applied to identify a linear function that can categorize data into two groups; this linear function was used as the decision boundary. The LRA is similar to the PLA, but it uses a sigmoid function as an activation function. The KNN algorithm was applied to monitor the position of data in the feature plane and then investigate the types of nearest $k$ neighbors. A final decision on Mira type was made based on the type with the largest number of neighbors. In this study, we set $k = 15$. The SVM algorithm was applied to identify a separating hyperplane for use as the decision boundary.

\begin{figure*}
    %\epsscale{1.10}
    \plotone{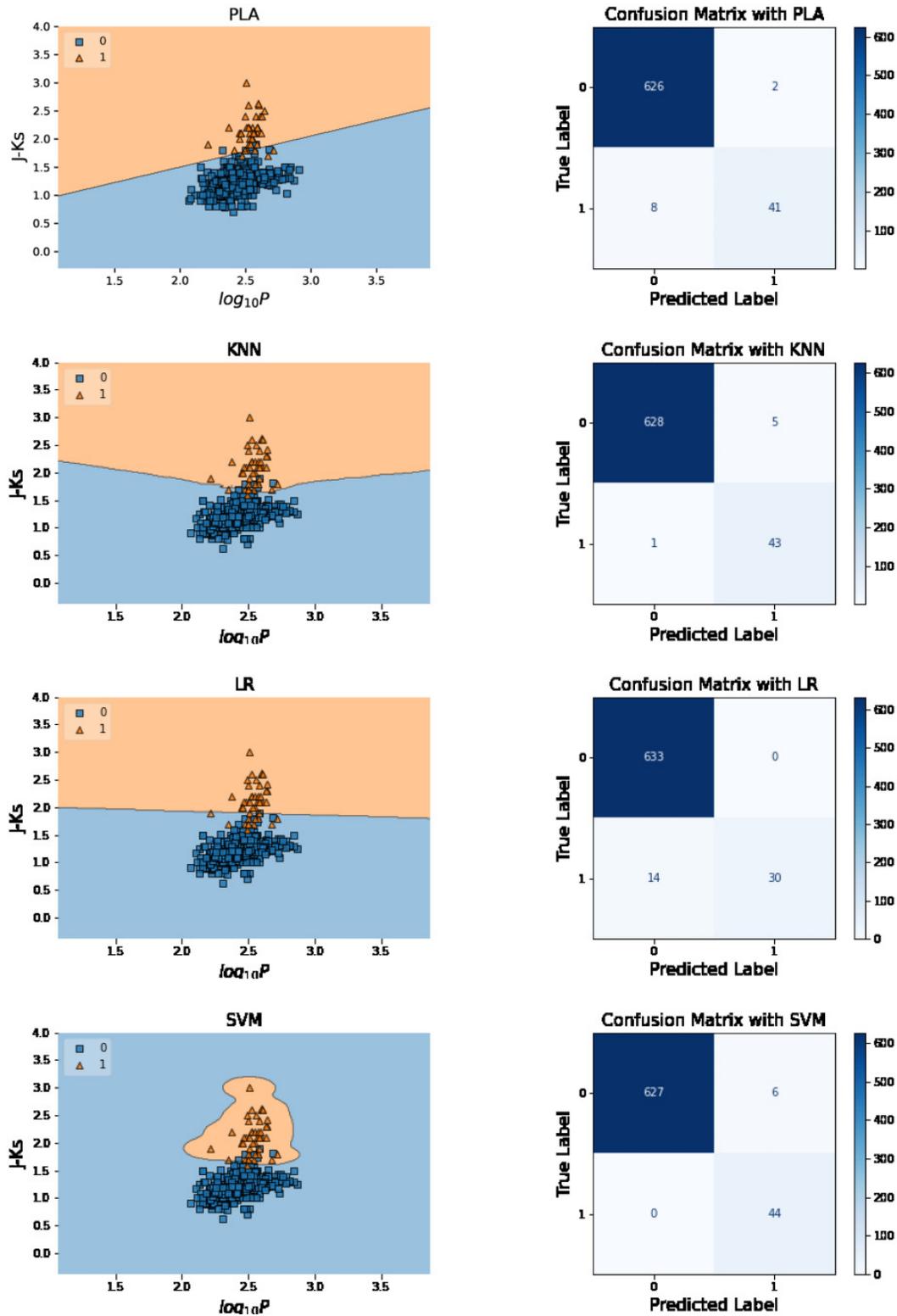}
    %\plottwo{Perceptron.eps}{confusion_matric_linear_model.eps}
    %\plottwo{LRM.eps}{confusion_matric_lR_model.eps}
    %\plottwo{KNN.eps}{confusion_matric_Knn_model.eps}
    %\plottwo{SVM.eps}{confusion_matric_SVM_model.eps}
    \caption{Left panel: Decision regions determined using the four ML algorithms. The orange sections show C-rich Miras, and the blue sections represent O-rich Miras. Right panel: Confusion matrices determined using each algorithm. The color bar represents the number of classified results.}
    \label{MLresult_new}
\end{figure*}

\begin{figure}
    \epsscale{1.15}
    \plotone{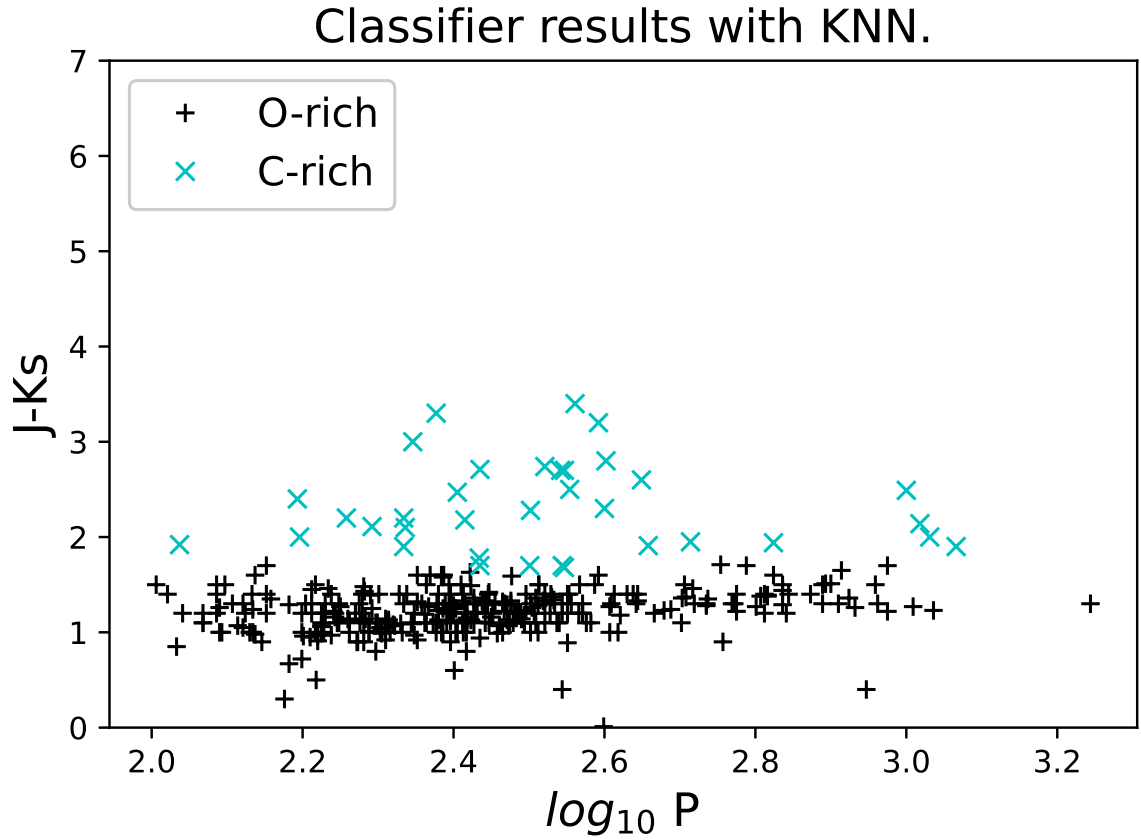}
    \caption{Results of the reclassification of unknown type of Miras given in \citet{2018AJ....156..112Y} using the KNN classifier. The black pluses and cyan crosses are for the O-rich and C-rich Miras, respectively.}
    \label{classify_result}
\end{figure}

In our study, 50\%  and 50\% of the Miras in the sample were used as training and test data, respectively.  We have tested these machine learning classifiers separately on the  LMC and M33 samples. In both samples the accuracy were greater than 95\%, however the decision boundaries were different. Hence, we adopted the M33 sample as the training data. The decision region obtained using each algorithms is presented in the left panels of Figure \ref{MLresult_new}, and the confusion matrix derived from the test data is displayed in the right panels of Figure \ref{MLresult_new}. All of the True--True ratios obtained from the test data were greater than $95\%$, except for those derived from the PLA; therefore, the relationship between the periods and $(J-K_s)$ colors could be applied to classify O-rich and C-rich Miras. We adopted the KNN algorithm to classify the 344 unclassified Miras presented in the Table 2 of \citet{2018AJ....156..112Y}. The number of O-rich and C-rich Miras was found to be 310 and 34, respectively; the corresponding results are presented in Figure \ref{classify_result} and Table \ref{datasummary}.

%This boundary had large margins. 

%\begin{figure*}
%    \epsscale{1.17}
%    \plottwo{Perceptron.eps}{KNN.eps}
%    \plottwo{LRM.eps}{SVM.eps}
%    \caption{Decision regions determined using each algorithm. The orange sections show C-rich Miras, and the blue 
%    sections represent O-rich Miras.}
%    \label{MLresult}
%\end{figure*}

%\begin{figure*}
%    \epsscale{1.17}
%    \plottwo{confusion_matric_linear_model.eps}{confusion_matric_Knn_model.eps}
%    \plottwo{confusion_matric_lR_model.eps}{confusion_matric_SVM_model.eps}
%    \caption{Confusion matrices determined using each algorithm. The color bar represents the number of classified results.}
%    \label{confustion}
%\end{figure*}

\begin{deluxetable*}{lcccccccccccccccc}
\label{datasummary}
\rotate
\tabletypesize{\footnotesize}
\tablecaption{Observed Properties of Miras in M33.}
\tablecolumns{17}
\tablewidth{0pt}
\tablehead{
	\colhead{ID} & \colhead{$P$ (days)} & \colhead{$\sigma_P$ (days)} & \colhead{Type}& \colhead{$\langle i \rangle$}& \colhead{$\sigma_i$} & \colhead{$i_{MAX}$}& \colhead{$\sigma_{imax}$} & \colhead{$\langle r \rangle$}& \colhead{$\sigma_r$} & \colhead{$r_{MAX}$}& \colhead{$\sigma_{rmax}$} & \colhead{$\langle g \rangle$}& \colhead{$\sigma_g$} & \colhead{$g_{MAX}$}& \colhead{$\sigma_{gmax}$} &\colhead{$E(B-V)$} 
          }
\startdata
01321450+3019349 & 260.7& 3.42 & O & 20.82 & 0.06 & 20.44 & 0.06 & 22.23 & 0.07 & 20.83 & 0.08 & 24.48 & 0.19 & 23.95 & 0.07 & 0.054 \\
01321654+3025260 & 304.1& 7.72 & O & 21.24 & 0.04 & 20.64 & 0.12 & 24.19 & 0.03 & 22.95 & 0.2 & 25.65 & 0.02 & 24.18 & 0.21 & 0.051 \\
01321897+3031226 & 255.4& 8.24 & O & 21.7 & 0.11 & 20.53 & 0.81 & 23.91 & 0.06 & 21.66 & 1.89 & 25.69 & 0.13 & 23.89 & 0.03 & 0.05 \\
01322179+3034063 & 356.0& 0.24 & O & 21.05 & 0.1 & 20.4 & 0.36 & 23.69 & 0.04 & 22.75 & 0.99 & 25.28 & 0.13 & 24.37 & 0.37 & 0.049 \\
01322351+3030590 & 252.3& 0.13 & O & 21.58 & 0.14 & 21.01 & 0.5 & 24.52 & 0.09 & 22.77 & 1.03 & 27.05 & 0.15 & 24.49 & 0.32 & 0.049\\
$\cdots$ &$\cdots$ &$\cdots$& $\cdots$ & $\cdots$ &$\cdots$ & $\cdots$ &$\cdots$ &$\cdots$ & $\cdots$ &$\cdots$ &$\cdots$ & $\cdots$ &$\cdots$ &$\cdots$ & $\cdots$ &$\cdots$ \\
\enddata
\tablecomments{The entire Table is published in its entirety in the machine-readable format. A portion is shown here for guidance regarding its form and content.}
\end{deluxetable*}
\section{The PL relation and Distance to M33}\label{sec4}

\begin{figure*}
    \epsscale{1.17}
    \plotone{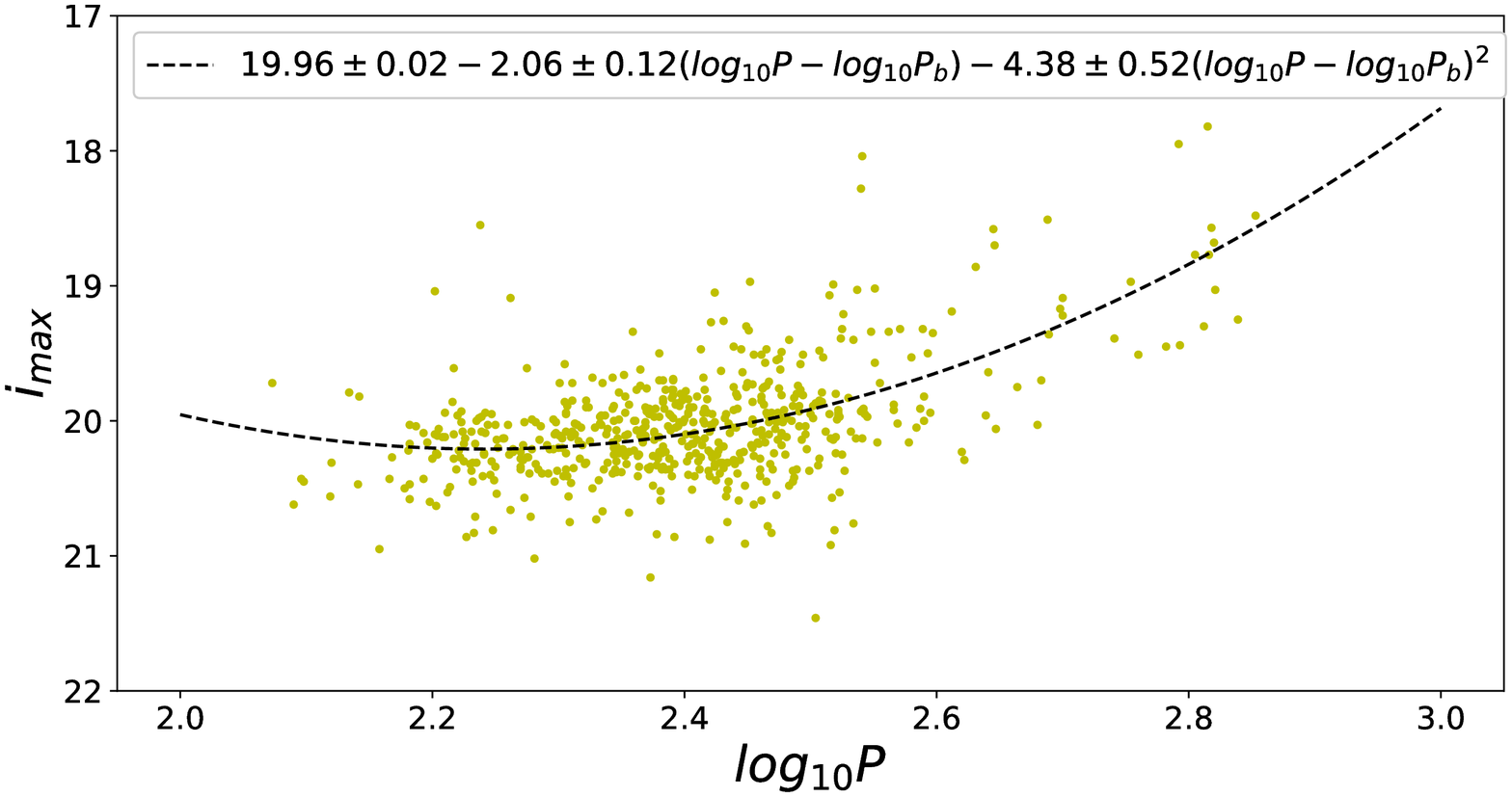}
    \plotone{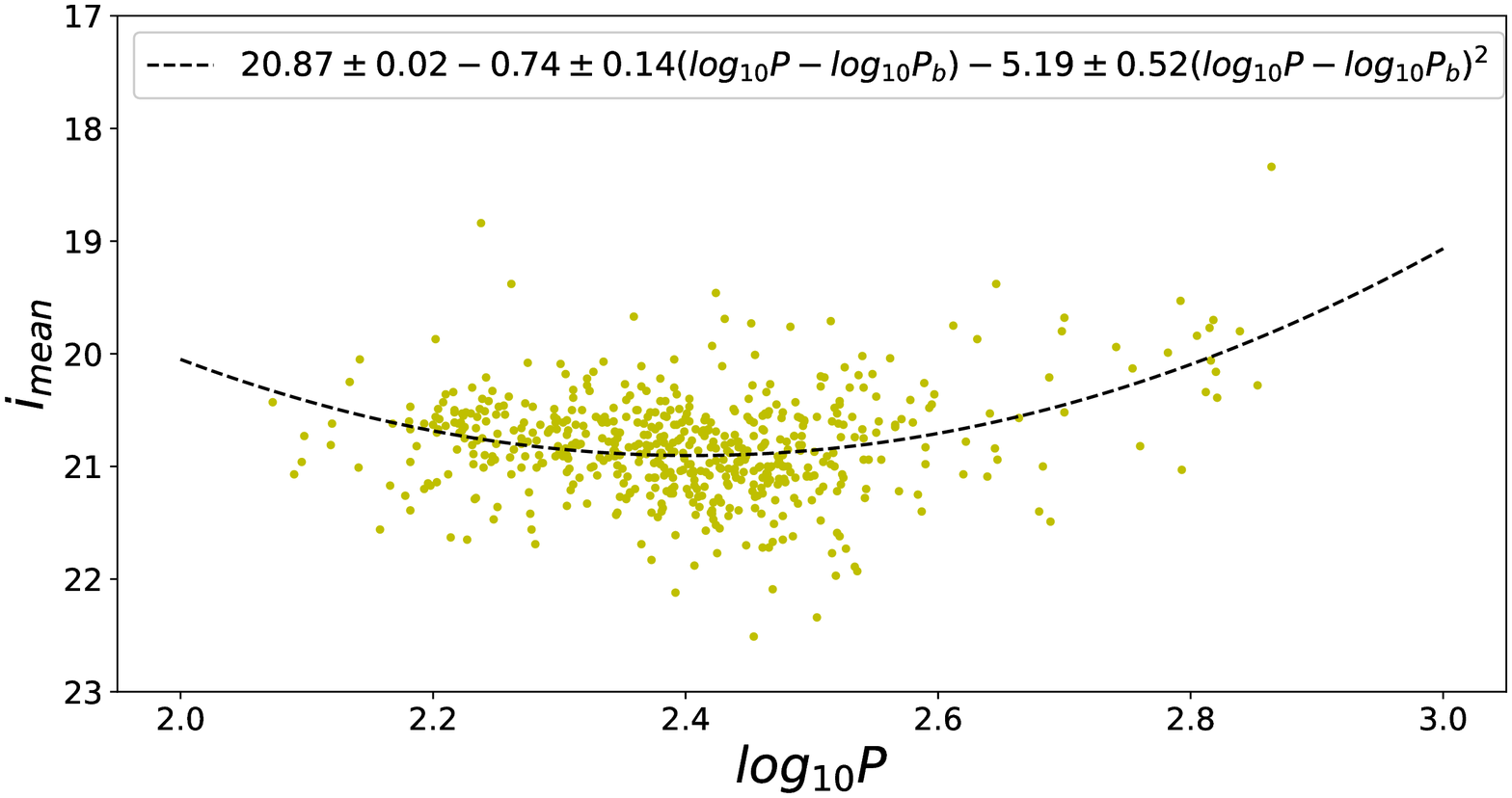}
    \caption{P--L relation for M33 O-rich Miras in the $i$-band at maximum (top panel) and mean (bottom panel) light. The dotted curves are the fitted quadratic P--L relation, i.e. equation (1), to the data, where the fitted coefficients are provided on the upper corner in each panels.}
    \label{PL-relaitoni}
\end{figure*}

Using the periods and magnitudes derived in Section \ref{sec31} and \ref{sec32}, distance to M33 can be determined by fitting the PL relation given in \citet{2022arXiv220300896O}. Before fitting the PL relation, we verified that the dispersion of PL relation is smaller at maximum light for the Miras in M33. We used a quadratic model to fit the P--L relation to the O-rich Mira $i$-band magnitudes at both maximum light and mean light (after corrected for extinction), where the quadratic model is given as:
\begin{equation}
    \label{quadratic}
    m = a + b_{1}(\log P-\log P_b) +b_{2}(\log P-\log P_b)^{2},
\end{equation}
\noindent with the break period adopted at $P_b = 300$~days \citep{2019ApJ...884...20B,2022arXiv220300896O}. The relevant results are presented in Figure \ref{PL-relaitoni}. The PL dispersion was found to be 0.42 at maximum light and 0.57 at mean light, confirming the earlier results.

%. Our results revealed that the maximum light dispersion from i-band data was smaller than the mean light dispersion from i-band data. 

\begin{figure*}
    \epsscale{1.17}
    \plotone{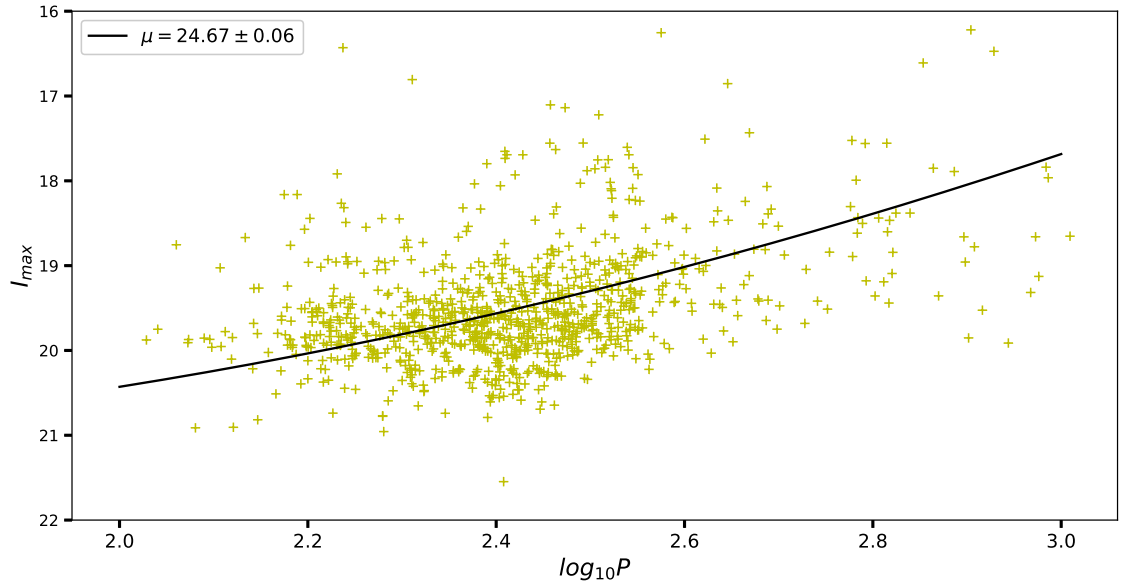}
    %\plotone{M33_PL_relation_mean.eps}
    \caption{Fitting of the $I$-band LMC PL relation (solid curves, calibrated with $\mu_{LMC} = 18.49$~mag) to the O-rich M33 Miras at maximum light. The $i$-band magnitudes at maximum light have been transformed to the $I$-band (see text for details) prior to the fittings. The fitted distance modulus for M33 is given on the upper-left corner.}
    %P--L relationship results obtained using transformed I-band data. The top panel shows the P--L relationship for maximum light time. The bottom panel shows the P--L relationship for mean light time. The yellow dots represent i bands and the line shows the P--L relationship function.}
    \label{PL-relaiton}
\end{figure*}

When fitting the P--L relation to derive the distance to M33, we only used the quadratic model of the P--L relation, based on the LMC O-rich Miras, taken from Table 2 of \citet{2022arXiv220300896O}, and fit to the O-rich Miras in M33 (including those reclassified in Section \ref{sec33}). A LMC distance modulus of $\mu_{LMC} = 18.49$~mag \citep{2019Natur.567..200P} was adopted when measuring the distance modulus of M33. Since the P--L relation presented in \citet{2022arXiv220300896O} is in the $I$-band, we transformed our $i$-band magnitudes, together with the corresponding $gr$-band magnitudes, at maximum light to $I$-band using the transformation given in \citet{2012ApJ...750...99T}.  We did not transform the $i$-band magnitudes at mean lights because the $gr$-band mean magnitudes are not reliable. The fitted P--L relation is presented in Figure \ref{PL-relaiton}, and we derived $\mu^{M33}_{max}=24.67\pm0.06$~mag at maximum light. Error on $\mu^{M33}_{max}$ is the quadrature sum of the errors on $\mu_{LMC}$ (0.048~mag), the estimated error in the fitted P--L relation (0.022~mag), and the dispersion from the photometric transformation (0.017~mag). 

Our derived distance modulus falls in the middle of the range of the distance modulus found in the literature (see Table \ref{Distancemodli}) and is the same as the recommended value of $24.67\pm0.07$~mag from \citet{2014AJ....148...17D}. \citet{2018AJ....156..112Y} found a distance modulus of 24.80~mag, albeit using a similar sample of O-rich Miras. We emphasize that both work used totally different datasets (such as observed in different filters) and methodologies (such as fitting of the P--L relations at mean and maximum light).  If we adopted the periods derived in \citet{2018AJ....156..112Y} and repeat the analysis, we found a distance modulus of $\mu^{M33}_{max} = 24.69 \pm 0.06$~mag, which is fully consistent with the distance modulus if using the periods we found in Section \ref{sec31}.

% Their $\Delta \mu$ was a combination of $\Delta a_0 $ from light curve fitting, correction for calculation mean magnitude $\Delta \bar{m} $, correction for differential extinction towards LMC and M33 $\Delta A_{\lambda} $ and correction for color terms at the mean color of O-rich Miras $\Delta C_t $. We have removed the extinction before fitting the P--L relation, and we only include the $\Delta a_0 $ in this study and in \citet{2022arXiv220300896O}.

%We adopted a distance of $49.59$~kpc or $\mu_0 = 18.49 \pm 0.046 $ \citep{2019Natur.567..200P} for LMC to measure the M33 distance modulus. Therefore, we transformed i-band data to determine the I-band magnitudes of each star with the color $ i - I$ \citep[the transformation data are given in][]{2006A&A...460..339J} in terms of maximum light time and mean light time. We used only LMC O-rich Miras from Table 2 \citep{2022arXiv220300896O} in the calculation. The  distance modulus $\mu_0$ derived at maximum light was $24.67 \pm 0.05,$ and that derived at mean light was $24.74 \pm 0.05$. The relevant results are presented in Figure \ref{PL-relaiton}. Different from those in the literature, the distance moduli in this study were between 24.52 and 24.80, as indicated in Table \ref{Distancemodli}. Therefore, both magnitude-related methods were applied to measured distance moduli.

\begin{deluxetable*}{ccccc}
\label{Distancemodli}
\tabletypesize{\footnotesize}
\tablecaption{Distance modulus for M33 from the literature.}
\tablecolumns{5}
\tablewidth{0pt}
\tablehead{
	\colhead{$\mu$} & \colhead{Method}& \colhead{Sample} & \colhead{Filter}& \colhead{Literature} 
 }
\startdata
$24.50\pm0.06$& TRGB  &  &g,i  &\citealt{2004MNRAS.350..243M} \\
$24.52\pm0.19$&Leavitt law & Cepheid &V,I &\citealt{2002ApJ...565..959L} \\
$24.53\pm0.11$& Leavitt law  & Cepheid  & B,V,Ic&\citealt{2009MNRAS.396.1287S}\\
$24.55\pm0.28$& Leavitt law  & Cepheid &V,I &\citealt{2007ApJ...671.1640A} \\ 
$24.57\pm0.06$& JAGB  &  & J,K &\citealt{2021ApJ...916...19Z}\\
$24.57\pm0.05$& TRGB  &  & i',g'& \citealt{2012ApJ...758...11C} \\
$24.58\pm0.10$&Leavitt law & Cepheid &V,I &\citealt{2001ApJ...553...47F} \\
$24.60\pm0.30$ & TRGB  &   &B,V,R,I &\citealt{1990AJ.....99..149W} \\
$24.62\pm0.06$& Leavitt law  & Cepheid & J,H,K &\citealt{2016AJ....151...88B}  \\
$24.62\pm0.07$& Leavitt law  & Cepheid & J,K,V,I &\citealt{2013ApJ...773...69G}  \\
$24.64\pm0.09$ &Leavitt law & Cepheid& B,V,R,I & \citealt{1991ApJ...372..455F} \\
$24.64\pm0.15$& TRGB   &  &  V,I &\citealt{2004AA...423..925G} \\
$24.67\pm0.05$& JAGB  &  & J,H,K &\citealt{2022ApJ...933..201L} \\
$24.67\pm0.06$&Leavitt law& Mira &I& {\bf This work}\\
$24.69\pm0.07$& TRGB  &  & V,I &\citealt{2004AJ....128..224T} \\
$24.70\pm0.11$& TRGB  &  & I &\citealt{2014AJ....148...17D} \\
$24.71\pm0.08$& JAGB  &  & J,H,K&\citealt{2022ApJ...933..201L}\\
$24.71\pm0.04$& TRGB &    &F555W,F606W,F814W  &\citealt{2007ApJ...661..815R} \\
$24.71\pm0.04$& Leavitt law & Cepheid & J,H,Ks &\citealt{2022ApJ...933..201L}\\
$24.72\pm0.14$& TRGB &   & V,I &\citealt{2004AJ....128..237B} \\
$24.72\pm0.11$& TRGB  &  &  J,H,K,I &\citealt{2022ApJ...933..201L} \\
$24.76\pm0.02$& Leavitt law  &  Cepheid & B,V,I  &\citealt{2011ApJS..193...26P}\\
$24.80\pm0.06$& Leavitt law& Mira & J,H,Ks &\citealt{2018AJ....156..112Y}\\
$24.80\pm0.10$ &Leavitt law & Cepheid& B,V,R & \citealt{1991MNRAS.250..438M} \\
$24.81\pm0.04$ & TRGB &   & V,I &\citealt{1993ApJ...417..553L} \\ 
$24.81_{-0.07}^{+0.19}$& TRGB  &  &F555W,F814W  &\citealt{2002AJ....123..244K} \\
24.82& TRGB  &  &V,I & \citealt{1997MNRAS.289..406S} \\
$24.84\pm0.10$& TRGB  &   &F814W,F606W  &\citealt{2009ApJ...704.1120U} \\
$24.93\pm0.18$& TRGB  &    & B,V,I &\citealt{2000ApJ...529..745F} \\
\enddata
\end{deluxetable*}

\section{Conclusion}\label{sec5}

In this work, we aimed to derive the distance to M33 using Mira variables. We started by re-determining the pulsating periods for 1378 Miras in M33 using available $gri$-band light curves. While performing the period analysis, we found that it is crucial to sample the full-amplitude light curve (in optical bands) rather than portions of the light curve around maximum light. This will be particularly important for Miras discovered in distant galaxies by the Vera C. Rubin Observatory Legacy Survey of Space and Time \cite[LSST,][]{lsst2019}, or other similar sky surveys, where it may not be possible to sample the light curves around minimum light due to limiting magnitudes.

In addition to period analysis, we showed that O-rich and C-rich Miras can be separated on $(J-K_s)$ PC plane using ML techniques, and hence we reclassified those Miras with unknown types in \citet{2018AJ....156..112Y}. Using all available O-rich M33 Miras, we demonstrated the P--L relation has a smaller dispersion at maximum light. Finally, we derived the distance modulus to M33 after transforming the photometry to $I$-band and fitted with the quadratic P--L relations given in \citet{2022arXiv220300896O}.  The derived distance modulus is $24.67 \pm 0.06$~mag using the P--L relations at maximum light, which is in good agreement with literature values. Our work demonstrated that both P--L relations at maximum and mean light for Miras can be used in distance scale measurements, this will be very useful in the era of LSST. 

\newpage

\begin{acknowledgments}
We thank an anonymous referee for the suggestions to improve the manuscript. We are thankful for funding from the Ministry of Science and Technology (MoST, Taiwan) under the contract 107-2119-M-008-014-MY2, 107-2119-M-008-012, 108-2628-M-007-005-RSP and 109-2112-M-008-014-MY3. AB acknowledges funding from the European Union’s Horizon 2020 research and innovation programme under the Marie Skłodowska-Curie grant agreement No. 886298.

Based on observations obtained with the 48-inch Samuel Oschin Telescope at the Palomar Observatory as part of the Zwicky Transient Facility project. ZTF is supported by the National Science Foundation under Grants No. AST-1440341 and AST-2034437 and a collaboration including current partners Caltech, IPAC, the Weizmann Institute of Science, the Oskar Klein Center at Stockholm University, the University of Maryland, Deutsches Elektronen-Synchrotron and Humboldt University, the TANGO Consortium of Taiwan, the University of Wisconsin at Milwaukee, Trinity College Dublin, Lawrence Livermore National Laboratories, IN2P3, University of Warwick, Ruhr University Bochum, Northwestern University and former partners the University of Washington, Los Alamos National Laboratories, and Lawrence Berkeley National Laboratories. Operations are conducted by COO, IPAC, and UW.

The Pan-STARRS1 Surveys (PS1) and the PS1 public science archive have been made possible through contributions by the Institute for Astronomy, the University of Hawaii, the Pan-STARRS Project Office, the Max-Planck Society and its participating institutes, the Max Planck Institute for Astronomy, Heidelberg and the Max Planck Institute for Extraterrestrial Physics, Garching, The Johns Hopkins University, Durham University, the University of Edinburgh, the Queen's University Belfast, the Harvard-Smithsonian Center for Astrophysics, the Las Cumbres Observatory Global Telescope Network Incorporated, the National Central University of Taiwan, the Space Telescope Science Institute, the National Aeronautics and Space Administration under Grant No. NNX08AR22G issued through the Planetary Science Division of the NASA Science Mission Directorate, the National Science Foundation Grant No. AST-1238877, the University of Maryland, Eotvos Lorand University (ELTE), the Los Alamos National Laboratory, and the Gordon and Betty Moore Foundation.
\end{acknowledgments}

\software{{\tt astropy} \citep{astropy2013,astropy2018}, {\tt gatspy} \citep{2015ApJ...812...18V}.}


\begin{thebibliography}{}
\bibitem[An et al.(2007)]{2007ApJ...671.1640A} An, D., Terndrup, D.~M., \& Pinsonneault, M.~H.\ 2007, \apj, 671, 1640. 
\bibitem[Astropy Collaboration et al.(2013)]{astropy2013} Astropy Collaboration, Robitaille, T.~P., Tollerud, E.~J., et al.\ 2013, \aap, 558, A33.

\bibitem[Astropy Collaboration et al.(2018)]{astropy2018} Astropy Collaboration, Price-Whelan, A.~M., Sip{\H{o}}cz, B.~M., et al.\ 2018, \aj, 156, 123.

\bibitem[Barsukova et al.(2011)]{2011MNRAS.413.1797B} Barsukova, E.~A., Goranskij, V.~P., Hornoch, K., et al.\ 2011, \mnras, 413, 1797.
\bibitem[Bellm et al.(2019)]{2019PASP..131a8002B} Bellm, E.~C., Kulkarni, S.~R., Graham, M.~J., et al.\ 2019, \pasp, 131, 018002.
\bibitem[Bhardwaj et al.(2016)]{2016AJ....151...88B} Bhardwaj, A., Kanbur, S.~M., Macri, L.~M., et al.\ 2016, \aj, 151, 88. 
\bibitem[Bhardwaj et al.(2019)]{2019ApJ...884...20B} Bhardwaj, A., Kanbur, S., He, S., et al.\ 2019, \apj, 884, 20.
\bibitem[Brooks et al.(2004)]{2004AJ....128..237B} Brooks, R.~S., Wilson, C.~D., \& Harris, W.~E.\ 2004, \aj, 128, 237.


\bibitem[Cardelli et al.(1989)]{1989ApJ...345..245C} Cardelli, J.~A., Clayton, G.~C., \& Mathis, J.~S.\ 1989, \apj, 345, 245
\bibitem[Cioni et al.(2001)]{2001A&A...377..945C} Cioni, M.-R.~L., Marquette, J.-B., Loup, C., et al.\ 2001, \aap, 377, 945.

\bibitem[Chambers et al.(2016)]{2016arXiv161205560C} Chambers, K.~C., Magnier, E.~A., Metcalfe, N., et al.\ 2016, arXiv:1612.05560.

\bibitem[Conn et al.(2012)]{2012ApJ...758...11C} Conn, A.~R., Ibata, R.~A., Lewis, G.~F., et al.\ 2012, \apj, 758, 11. 

\bibitem[Dekany et al.(2020)]{dec20} Dekany, R., Smith, R.~M., Riddle, R., et al.\ 2020, \pasp, 132, 038001
\bibitem[de Grijs \& Bono(2014)]{2014AJ....148...17D} de Grijs, R. \& Bono, G.\ 2014, \aj, 148, 17. 

\bibitem[Feast(1984)]{1984MNRAS.211P..51F} Feast, M.~W.\ 1984, \mnras, 211, 51P 
\bibitem[Feast et al.(1989)]{1989MNRAS.241..375F} Feast, M.~W., Glass, I.~S., Whitelock, P.~A., et al.\ 1989, \mnras, 241, 375 
\bibitem[Ferrarese et al.(2000)]{2000ApJ...529..745F} Ferrarese, L., Mould, J.~R., Kennicutt, R.~C., et al.\ 2000, \apj, 529, 745. 

\bibitem[Freedman et al.(1991)]{1991ApJ...372..455F} Freedman, W.~L., Wilson, C.~D., \& Madore, B.~F.\ 1991, \apj, 372, 455. 
\bibitem[Freedman et al.(2001)]{2001ApJ...553...47F} Freedman, W.~L., Madore, B.~F., Gibson, B.~K., et al.\ 2001, \apj, 553, 47. 


\bibitem[Galleti et al.(2004)]{2004AA...423..925G} Galleti, S., Bellazzini, M., \& Ferraro, F.~R.\ 2004, \aap, 423, 925. 
\bibitem[Gieren et al.(2013)]{2013ApJ...773...69G} Gieren, W., G{\'o}rski, M., Pietrzy{\'n}ski, G., et al.\ 2013, \apj, 773, 69. 

\bibitem[Graham et al.(2019)]{2019PASP..131g8001G} Graham, M.~J., Kulkarni, S.~R., Bellm, E.~C., et al.\ 2019, \pasp, 131, 078001. 

\bibitem[Green et al.(2018)]{2018MNRAS.478..651G} Green, G.~M., Schlafly, E.~F., Finkbeiner, D., et al.\ 2018, \mnras, 478, 651. 

\bibitem[Glass \& Evans(1981)]{1981Natur.291..303G} Glass, I.~S. \& Evans, T.~L.\ 1981, \nat, 291, 303. 

\bibitem[Hartman et al.(2006)]{2006MNRAS.371.1405H} Hartman, J.~D., Bersier, D., Stanek, K.~Z., et al.\ 2006, \mnras, 371, 1405.

\bibitem[Ivezi{\'c} et al.(2019)]{lsst2019} Ivezi{\'c}, {\v{Z}}., Kahn, S.~M., Tyson, J.~A., et al.\ 2019, \apj, 873, 111.

\bibitem[Iwanek et al.(2021)]{2021ApJ...919...99I} Iwanek, P., Soszy{\'n}ski, I., \& Koz{\l}owski, S.\ 2021, \apj, 919, 99. 
\bibitem[Iwanek et al.(2021)]{2021ApJS..257...23I} Iwanek, P., Koz{\l}owski, S., Gromadzki, M., et al.\ 2021, \apjs, 257, 23. 


\bibitem[Kaiser et al.(2010)]{2010SPIE.7733E..0EK} Kaiser, N., Burgett, W., Chambers, K., et al.\ 2010, \procspie, 7733, 77330E.
\bibitem[Kanbur et al.(1997)]{1997MNRAS.289..428K} Kanbur, S.~M., Hendry, M.~A., \& Clarke, D.\ 1997, \mnras, 289, 428
\bibitem[Kim et al.(2002)]{2002AJ....123..244K} Kim, M., Kim, E., Lee, M.~G., et al.\ 2002, \aj, 123, 244. 

\bibitem[Kulkarni(2013)]{2013ATel.4807....1K} Kulkarni, S.~R.\ 2013, The Astronomer's Telegram, 4807

\bibitem[Lafler \& Kinman(1965)]{1965ApJS...11..216L} Lafler, J. \& Kinman, T.~D.\ 1965, \apjs, 11, 216.

\bibitem[Law et al.(2009)]{2009PASP..121.1395L} Law, N.~M., Kulkarni, S.~R., Dekany, R.~G., et al.\ 2009, \pasp, 121, 1395.
%\bibitem[Leavitt \& Pickering(1912)]{1912HarCi.173....1L} Leavitt, H.~S. \& Pickering, E.~C.\ 1912, Harvard College Observatory Circular, 173
\bibitem[Lebzelter et al.(2018)]{2018AA...616L..13L} Lebzelter, T., Mowlavi, N., Marigo, P., et al.\ 2018, \aap, 616, L13.

\bibitem[Lee et al.(1993)]{1993ApJ...417..553L} Lee, M.~G., Freedman, W.~L., \& Madore, B.~F.\ 1993, \apj, 417, 553. 
\bibitem[Lee et al.(2002)]{2002ApJ...565..959L} Lee, M.~G., Kim, M., Sarajedini, A., et al.\ 2002, \apj, 565, 959.
\bibitem[Lee et al.(2022)]{2022ApJ...933..201L} Lee, A.~J., Rousseau-Nepton, L., Freedman, W.~L., et al.\ 2022, \apj, 933, 201. 
\bibitem[Lee et al.(2021)]{2021ApJ...911...51L} Lee, C.-D., Ou, J.-Y., Yu, P.-C., et al.\ 2021, \apj, 911, 51.


%\bibitem[Lomb(1976)]{1976Ap&SS..39..447L} Lomb, N.~R.\ 1976, \apss, 39, 447.


\bibitem[Masci et al.(2019)]{2019PASP..131a8003M} Masci, F.~J., Laher, R.~R., Rusholme, B., et al.\ 2019, \pasp, 131, 018003.
\bibitem[McConnachie et al.(2004)]{2004MNRAS.350..243M} McConnachie, A.~W., Irwin, M.~J., Ferguson, A.~M.~N., et al.\ 2004, \mnras, 350, 243.

\bibitem[Merrill(1960)]{1960ApJ...131..385M} Merrill, P.~W.\ 1960, \apj, 131, 385. 
\bibitem[Metcalfe \& Shanks(1991)]{1991MNRAS.250..438M} Metcalfe, N. \& Shanks, T.\ 1991, \mnras, 250, 438. 

\bibitem[Ou \& Ngeow(2022)]{2022arXiv220300896O} Ou, J.-Y. \& Ngeow, C.-C.\ 2022, \aj, 163, 192.

\bibitem[Pellerin \& Macri(2011)]{2011ApJS..193...26P} Pellerin, A. \& Macri, L.~M.\ 2011, \apjs, 193, 26. 
\bibitem[Pietrzy{\'n}ski et al.(2019)]{2019Natur.567..200P} Pietrzy{\'n}ski, G., Graczyk, D., Gallenne, A., et al.\ 2019, \nat, 567, 200 

\bibitem[Rau et al.(2009)]{2009PASP..121.1334R} Rau, A., Kulkarni, S.~R., Law, N.~M., et al.\ 2009, \pasp, 121, 1334.

\bibitem[Riebel et al.(2010)]{2010ApJ...723.1195R} Riebel, D., Meixner, M., Fraser, O., et al.\ 2010, \apj, 723, 1195.
\bibitem[Rizzi et al.(2007)]{2007ApJ...661..815R} Rizzi, L., Tully, R.~B., Makarov, D., et al.\ 2007, \apj, 661, 815. 

\bibitem[Salaris \& Cassisi(1997)]{1997MNRAS.289..406S} Salaris, M. \& Cassisi, S.\ 1997, \mnras, 289, 406. 
%\bibitem[Scargle(1982)]{1982ApJ...263..835S} Scargle, J.~D.\ 1982, \apj, 263, 835.
\bibitem[Scowcroft et al.(2009)]{2009MNRAS.396.1287S} Scowcroft, V., Bersier, D., Mould, J.~R., et al.\ 2009, \mnras, 396, 1287. 
\bibitem[Soszy{\'n}ski et al.(2005)]{2005AcA....55..331S} Soszy{\'n}ski, I., Udalski, A., Kubiak, M., et al.\ 2005, \actaa, 55, 331

\bibitem[Soszy{\'n}ski et al.(2009)]{2009AcA....59..239S} Soszy{\'n}ski, I., Udalski, A., Szyma{\'n}ski, M.~K., et al.\ 2009, \actaa, 59, 239

\bibitem[Tiede et al.(2004)]{2004AJ....128..224T} Tiede, G.~P., Sarajedini, A., \& Barker, M.~K.\ 2004, \aj, 128, 224. 
\bibitem[Tonry et al.(2012)]{2012ApJ...750...99T} Tonry, J.~L., Stubbs, C.~W., Lykke, K.~R., et al.\ 2012, \apj, 750, 99.

\bibitem[U et al.(2009)]{2009ApJ...704.1120U} U, V., Urbaneja, M.~A., Kudritzki, R.-P., et al.\ 2009, \apj, 704, 1120.


\bibitem[VanderPlas \& Ivezi{\'c}(2015)]{2015ApJ...812...18V} VanderPlas, J.~T. \& Ivezi{\'c}, {\v{Z}}.\ 2015, \apj, 812, 18.


\bibitem[Wang et al.(2022)]{2022ApJS..260...41W} Wang, Y., Gao, J., Ren, Y., et al.\ 2022, \apjs, 260, 41. 

\bibitem[Wilson et al.(1990)]{1990AJ.....99..149W} Wilson, C.~D., Freedman, W.~L., \& Madore, B.~F.\ 1990, \aj, 99, 149. 
\bibitem[Whitelock et al.(2008)]{2008MNRAS.386..313W} Whitelock, P.~A., Feast, M.~W., \& Van Leeuwen, F.\ 2008, \mnras, 386, 313. 
\bibitem[Whitelock(2012)]{2012ApSS.341..123W} Whitelock, P.~A.\ 2012, \apss, 341, 123.


\bibitem[Yuan et al.(2017)]{2017AJ....153..170Y} Yuan, W., He, S., Macri, L.~M., et al.\ 2017, \aj, 153, 170.
\bibitem[Yuan et al.(2017)]{2017AJ.154..149Y} Yuan, W., Macri, L.~M., He, S., et al.\ 2017, \aj, 154, 149. 
\bibitem[Yuan et al.(2018)]{2018AJ....156..112Y} Yuan, W., Macri, L.~M., Javadi, A., et al.\ 2018, \aj, 156, 112.
\bibitem[Zgirski et al.(2021)]{2021ApJ...916...19Z} Zgirski, B., Pietrzy{\'n}ski, G., Gieren, W., et al.\ 2021, \apj, 916, 19. 




\end{thebibliography}
\end{document}